\documentclass[10pt, conference]{IEEEtran}
\usepackage{cite}
\usepackage{amsmath,amssymb,amsfonts}
\usepackage{algorithmic}
\usepackage{graphicx}
\usepackage{textcomp}
\usepackage{url}
\usepackage{balance}

\usepackage{enumitem}
\usepackage{xcolor}
\usepackage{booktabs}

\usepackage{listings}
\usepackage[most]{tcolorbox}

\makeatletter
\let\MYcaption\@makecaption
\makeatother

\makeatletter
\let\@makecaption\MYcaption
\makeatother

\usepackage{caption}
\usepackage{subcaption}
\usepackage{float}

\lstdefinelanguage{COBOL}{
    morecomment=[l]{*},
    morestring=[b]",
    sensitive=false,
    alsoletter={-},
}

\tcbset{
    mypromptbox/.style={
        colback=gray!10,    
        colframe=gray!70,   
        fonttitle=\bfseries,
        fontupper=\ttfamily,
        coltitle=black,
        boxrule=0.5mm,
        boxsep=5pt,
        left=5pt,
        right=5pt,
        top=5pt,
        bottom=5pt
    }
}

\title{Quality Evaluation of COBOL to Java Code Transformation}

\author{
\IEEEauthorblockN{Shmulik Froimovich, Raviv Gal, Wesam Ibraheem, Avi Ziv}
\IEEEauthorblockA{
Quality Technologies Department\\
IBM Research - Israel\\
Haifa, Israel\\
Email: shmulik.froimovich@ibm.com, \{ravivg, wesam, aziv\}@il.ibm.com
}
}

\begin{document}

\maketitle
\begin{abstract}
We present an automated evaluation system for assessing COBOL-to-Java code translation within IBM’s watsonx Code Assistant for Z (WCA4Z). The system addresses key challenges in evaluating LLM-based translators, including model opacity and the complexity of translation quality assessment. Our approach combines analytic checkers with LLM-as-a-judge (LaaJ) techniques to deliver scalable, multi-faceted evaluations. The system supports continuous integration workflows, enables large-scale benchmarking, and reduces reliance on manual review. We describe the system architecture, evaluation strategies, and reporting mechanisms that provide actionable insights for developers and project managers, facilitating the evolution of high-quality, modernized codebases.
\end{abstract}


\section{Introduction}
\label{introduction-section}

Artificial Intelligence (AI), and large language models (LLMs) in particular, have revolutionized the software development process by introducing AI-powered services such as code generation and explanation into nearly all modern development environments. These services enhance code quality and reduce development time and effort.

\textit{IBM watsonx Code Assistant for Z} (WCA4Z)~\cite{wca4z} is IBM’s code assistant tailored specifically for software development on IBM mainframe platforms. WCA4Z offers standard AI capabilities found in other code assistants, such as program analysis, code generation and explanation, and optimization. What sets WCA4Z apart is its support for programming languages unique to the mainframe environment, including COBOL, PL/I, and JCL. To support these languages, WCA4Z employs LLMs fine-tuned for its specific tasks and supported languages. Another distinctive feature of WCA4Z is its strong support for application modernization. This includes automatic refactoring of monolithic applications and code transformation from traditional mainframe languages, such as COBOL, to modern languages like Java. 

This paper presents the system we developed to evaluate the quality of the COBOL-to-Java translation subsystem. Broadly, the evaluation system collects results from benchmark runs, processes them, assesses the quality of each translation, and analyzes the results to provide actionable insights to its users. It is important to note that the creation of the benchmarks used in the evaluation process is outside the scope of this system and paper.

There are two primary challenges in evaluating LLM-based code translators. The first stems from the nature of LLMs themselves. These models are black boxes, offering no insight into their internal reasoning or why a particular output is generated for a given input. Moreover, LLMs exhibit traits such as hallucinations and occasional lack of robustness, which complicate evaluation. As a result, evaluation must rely solely on input-output behavior while accounting for these characteristics. The second challenge lies in assessing the quality of code translation. Evaluating the correctness of a COBOL-to-Java translation is inherently difficult, as proving equivalence between two programs is undecidable~\cite{sipser2012introduction}. This challenge is magnified considering the large semantic difference between the source and target languages, namely, COBOL and Java.

This paper makes three key contributions. First, we describe the architecture and operation of the evaluation system. The system is a data-driven pipeline that ingests translation results from the code transformation component, processes and stores them in a database, and then invokes a suite of evaluators and checkers to assess various aspects of the translations. These evaluation results are also stored and used to analyze translator performance, providing insights to stakeholders such as project managers and model developers. Notably, the evaluation system is built on a shared infrastructure also used by other LLM-based components in WCA4Z, such as code explanation and generation.

Second, we detail our approach to evaluating individual translations. As there is no single definitive method for assessing COBOL-to-Java translation quality, we employ a diverse set of checkers and evaluators, each targeting specific aspects of the translation. These range from simple syntactic checks (e.g., whether the translated code is parsable), to semantic checks (e.g., correctness of SQL operation translations), to full compilation and execution of the translated code.

A complementary approach involves using LLMs as judges (LaaJs) to assess translation quality~\cite{lagakis2024evaai}. LaaJs can rate translation quality and identify issues, offering a holistic evaluation. However, they require domain-specific expertise in COBOL-to-Java translation, which can limit their effectiveness. We find that combining precise but partial analytic checkers with comprehensive but less precise LaaJs yields a balanced and informative evaluation.

Third, we describe the analysis and reporting component, which synthesizes evaluation results into actionable insights. For example, it aggregates translation evaluations for a given version to provide an overall quality assessment and compare different versions. It can also identify specific translation issues, such as COBOL statements that frequently result in unparsable Java code.

The evaluation system has been used throughout the development of WCA4Z’s code transformation component, from early prototypes to its current, rapidly maturing state. The pipeline enables large-scale evaluation, facilitating experimentation and improvement of the translation subsystem. Our hybrid evaluation approach, combining analytic checkers and LaaJs, has significantly reduced the need for manual \textit{Subject-Matter Expert} (SME) involvement in the evaluation process.

The remainder of this paper is organized as follows: Section~\ref{wca4z-sec} provides a high-level overview of watsonx Code Assistant for Z and its COBOL-to-Java transformation component. Section~\ref{wca4z-eval-sec} describes our evaluation framework for the LLM-based translation subsystem. Section~\ref{single-trans-sec} details the evaluation of individual translations. Section~\ref{reporting-sec} discusses how evaluation results are analyzed and presented to development and project management teams. Finally, Section~\ref{summary-sec} concludes the paper.

\section{Code Transformation in IBM watsonx Code Assistant for Z  (WCA4Z)}
\label{wca4z-sec}

Mainframes are widely used in industries like banking, insurance, and government for mission-critical applications due to their reliability, scalability, and security. Despite the rise of cloud computing, mainframes remain essential for processing large-scale workloads and maintaining legacy systems. 

COBOL (Common Business-Oriented Language)~\cite{stern2013cobol} is a high-level programming language developed in the late 1950s, designed specifically for business, finance, and administrative systems. It emphasizes readability and uses English-like syntax, making it accessible for non-technical stakeholders. 

COBOL and mainframes are deeply interconnected, as COBOL was specifically designed to run efficiently on mainframe systems. It is estimated that over 200 billion lines of COBOL code are currently live and operational on mainframes~\cite{cobol-loc}. These applications are owned by a highly regulated industry, making the language and platform a foundational pair in industries that rely on high-volume, reliable transaction processing.

Modernizing mainframe applications is a critical challenge for enterprises relying on IBM Z systems. These systems often run mission-critical workloads written in legacy languages like COBOL and PL/I, which are increasingly difficult to maintain due to a shrinking pool of skilled developers. IBM watsonx Code Assistant for Z (WCA4Z)~\cite{wca4z} addresses this challenge by leveraging generative AI and automation to accelerate the application modernization lifecycle.

\begin{figure}[t]
\centering
\includegraphics[width=\columnwidth]{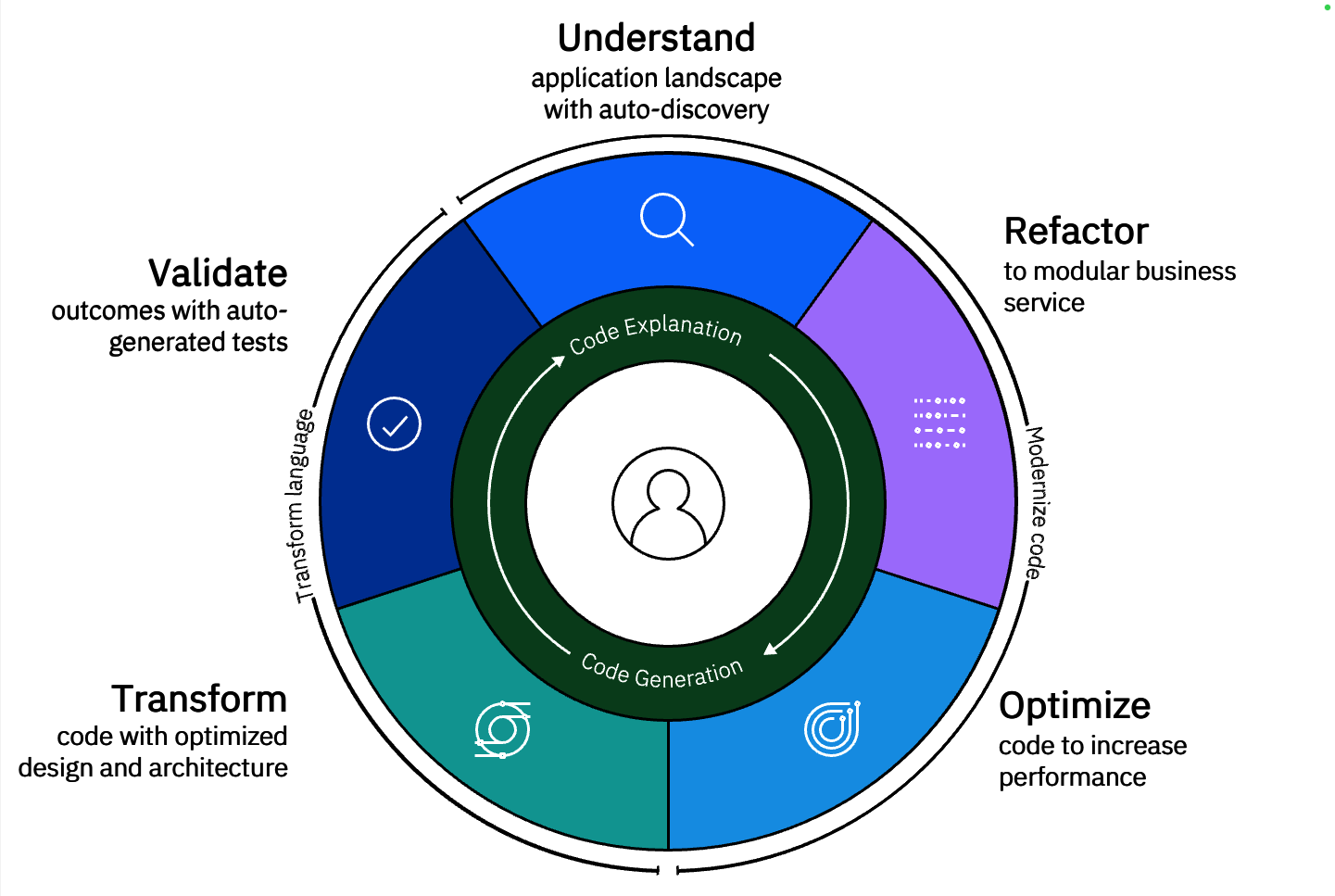}
\caption{Architecture diagram of IBM watsonx Code Assistant for Z (WCA4Z)}
\label{fig:wca4z_architecture}
\end{figure}

WCA4Z supports a comprehensive modernization process, illustrated in Figure~\ref{fig:wca4z_architecture},  that begins with application discovery. Using tools like Application Discovery and Delivery Intelligence (ADDI)~\cite{addi}, the system automatically analyzes source libraries to generate accurate call graphs and dependency maps. Once the structure and dependencies are understood, the system helps the user to decide the refactoring strategy. Here, the user can select a functionality (e.g., onboarding a new customer service). Using backward and forward slicing techniques, WCA4Z identifies and extracts modular business services from monolithic COBOL or PL/I applications. Now the user can decide whether to keep the new service in the legacy program language or transform it to Java. 

The code transformation phase in WCA4Z leverages a fine-tuned large language model (LLM) to convert COBOL code into optimized, object-oriented Java. Unlike traditional tools that perform line-by-line translations---often resulting in what is colloquially referred to as ``JOBOL''~\cite{jobol} (COBOL code written in Java syntax)---WCA4Z adopts a two-phase, semantically driven approach.
In the first phase, called {\em Class Designer} (CD), the entire COBOL program is analyzed to generate a proposed Java class design. This design includes the overall structure and class relationships, allowing the user to review and refine the architecture before proceeding. Once the design is accepted, WCA4Z generates the Java class files with method headers in place, but without the internal logic.

The second phase focuses on method-level transformation. Developers can navigate through each method, examine the corresponding COBOL paragraph, and request a translation into Java. The generated code can then be reviewed, edited if necessary, and approved before being inserted into the class file. This human-in-the-loop process ensures that each transformation step is both accurate and aligned with the intended design, combining the strengths of automation with expert oversight.

To ensure performance, WCA4Z provides insights into inefficiencies at the source code level, enabling targeted improvements. Finally, the system generates unit tests to validate the semantic equivalence of the transformed Java code with the original COBOL, reducing risk and increasing confidence in the modernization process.

A key innovation in WCA4Z is its use of LLMs fine-tuned for mainframe languages and use cases. These models enable high-quality code generation, explanation, and transformation, significantly reducing the manual effort and expertise traditionally required for mainframe modernization.

We use the running example in Figure~\ref{fig:running-ex} throughout the paper to illustrate both the translation process of a single COBOL paragraph and the quality evaluation of the translation. The example is taken from the {\em general insurance application} (gennApp)~\cite{genapp}, a demo CICS (Customer Information Control System)~\cite{cics} application. Note that we edited both the source COBOL code and the translated Java code to better fit the paper format and help illustrating some aspects of the quality evaluation.

Figure~\ref{fig:ex-cobol-src} provides the source code of the paragraph. This COBOL code executes a CICS transaction to write customer data to a file named \texttt{'KSDSCUST'}. If the write transaction is unsuccessful, it sets the return code to '80', writes an error message, and abnormally ends the program. 

\begin{figure*}[t]
    \centering
    \begin{subfigure}[b]{0.75\columnwidth}
    \begin{lstlisting}[language=COBOL]
MAINLINE SECTION.

     Move EIBCALEN To WS-Commarea-Len.

     Exec CICS Write File('KSDSCUST')
               From(CA-Customer-Num)
               Length(WS-Commarea-Len)
               Ridfld(CA-Customer-Num)
               KeyLength(10)
               RESP(WS-RESP)
     End-Exec.

     If WS-RESP Not = DFHRESP(NORMAL)
       Move EIBRESP2 To WS-RESP2
       MOVE '80' TO CA-RETURN-CODE
       PERFORM WRITE-ERROR-MESSAGE

       EXEC CICS ABEND 
                 ABCODE('LGV0') NODUMP 
       END-EXEC

       EXEC CICS RETURN END-EXEC
     End-If.
    \end{lstlisting}
    \caption{Source COBOL code}
    \label{fig:ex-cobol-src}
    \end{subfigure}
    \hfill
    \begin{subfigure}[b]{1.20\columnwidth}
    \centering
    \centering
    \begin{lstlisting}[language=Java]
public void invokeMainline(Dfhcommarea dfhcommarea) {
    int wsResp;
    int wsResp2;
    int wsCommareaLen = Task.getTask().getApplicationContext().getPlatform().getCommAreaLen();
    try {
        KeyedFile jdeclKeyedFile = new KeyedFile();
        jdeclKeyedFile.setName("KSDSCUST");
        RecordHolder jdeclRecordHolder = new RecordHolder();
        String jdeclLocalCcsid = System.getProperty("com.ibm.cics.jvmserver.local.ccsid");
        Charset jdeclLocalCharSet = Charset.forName(jdeclLocalCcsid);
        jdeclKeyedFile.write(caCustomerNum, wsCommareaLen, jdeclRecordHolder);
    } catch (CicsException e) {
        wsResp = e.getRESP();
        wsResp2 = e.getRESP2();
        dfhcommarea.setCaReturnCode(80);
        mainlineWriteErrorMessage(dfhcommarea, wsResp, wsResp2);
        Task.getTask().abend("LGV0", true);
        return;
    }
}
    \end{lstlisting}
    \caption{Translated Java code}
    \label{fig:ex-java-code}
    \end{subfigure}

    \begin{subfigure}[t]{0.98\textwidth}
    \begin{lstlisting}
## Variable Map:   getter                           setter
WS-RESP            wsResp                           wsResp = val
WS-RESP2           wsResp2                          wsResp2 = val
WS-Commarea-Len    wsCommareaLen                    wsCommareaLen = val
CA-RETURN-CODE     dfhcommarea.getCaReturnCode()    dfhcommarea.setCaReturnCode(val)
CA-CUSTOMER-NUM    dfhcommarea.getCaCustomerNum()   dfhcommarea.setCaCustomerNum(val)

## Method Map:
WRITE-ERROR-MESSAGE          mainlineWriteErrorMessage(dfhcommarea, wsResp, wsResp2)

## Class Map:
public void invokeMainline(Dfhcommarea dfhcommarea){
    int wsCommareaLen;
    int wsResp;
    int wsResp2;
}
    \end{lstlisting}
    \caption{Additional translation information}
    \label{fig:ex-var-map}
    \end{subfigure} 
\caption{COBOL to Java translation example}
\label{fig:running-ex}
\end{figure*} 

In addition to the source COBOL, the LLM also receives additional information needed for the translation. This information, created by the Class Designer, is given in Figure~\ref{fig:ex-var-map}. It contains the variable mappings between COBOL variables to Java variables. For example, \texttt{WS-RESP2} COBOL variables should be translated to local variable \texttt{wsResp2} and \texttt{CA-CUSTOMER-NUM}, which is part of the \texttt{DFHCOMMAREA} record that is used for communication area between CICS programs, is translated to getter and setter method in the \texttt{Dfhcommarea} data class. The additional information also contains the class map that includes the signature of generated Java method and some local variables it must declare.

The resulting translated Java code is given in Figure~\ref{fig:ex-java-code}. It shows some of the differences between the COBOL and Java code. First is the setup needed for the CICS objects in lines 7--12 of the Java code, which are hidden in the \texttt{EXEC CICS} COBOL statement. In addition, while the COBOL code checks a returned status variable \texttt{WS-RESP} for failure, Java handles this with exception.

\section{WCA4Z Evaluation Overview}
\label{wca4z-eval-sec}

This section presents the evaluation framework for the translation subsystem of the WCA4Z code transformation component. Specifically, it focuses on the part of the system that receives a prompt containing a COBOL paragraph and supplementary information from the Class Designer, and produces a translated Java method as output from the LLM, following postprocessing.

The evaluation system is designed to meet industry-grade quality standards, which significantly influenced its requirements and specification. The WCA4Z development team defined two key requirements at the outset. First, the evaluation should be conducted at the transformation component level, not just the translation subsystem level. In addition, the transformation component must remain decoupled from the evaluation system.

To support these requirements, the team developed a testing driver that replaces the IDE and interactes with the transformation component. This driver executes tests, collects results along with all necessary metadata, and packages them into a \texttt{.jsonl} file (a JSON Lines file, where each line represents a single test case). Communication between the driver and the evaluation system is handled via a shared Git repository: the driver pushes test result files, and the evaluation system automatically consumes them.

The demand for industry-grade evaluation necessitates large-scale, frequently executed regression tests. This, in turn, requires full automation of both test execution and evaluation. The testing driver addresses the automation of test execution, while the evaluation platform is designed to support fully automated analysis.

The evaluation is based on static tests, enabling consistent comparisons across different versions of the system, both at the aggregate and individual test levels. The benchmarking team developed several datasets (or benchmarks), each corresponding to a specific application (e.g., GenApp) or targeting particular COBOL language features (e.g., basic COBOL or CICS). Each dataset includes multiple programs, with selected paragraphs marked for translation. These marked paragraphs are referred to as \textit{datapoints}.

To evaluate dataset quality and support analysis, we apply a hierarchical coverage model to the input COBOL. This model has three levels: The first level, \textit{category}, split the COBOL statements to large topics, such as basic COBOL or SQL. The second level, \textit{subcategory}, has elements for major parts of the categories. For example, each basic COBOL statement is a subcategory in the basic COBOL category and each CICS operation is a subcategory in the CICS category. The lowest level, \textit{sub-subcategory}, has elements for certain aspects of the subcategory. For example, IF statement has sub-subcategories for if with else clause, nested if, and if with complex condition. Since the coverage evaluation is based on static tests and coverage is defined on the input, we measure coverage for new datapoints as they are introduced into the system.

The evaluation system is a structured, data-centric pipeline designed to ensure traceability, reproducibility, and longitudinal comparison across versions. At its core is a relational database that persistently stores all evaluation data and supports the evaluation tools, analysis modules, and reporting engines. The database schema reflects the evaluation data model, which consists of four main sections:

\begin{description}
    \item[Static Data:] Includes datasets, datapoints, and auxiliary information such as the source programs.
    \item \textbf{Evaluation Sets and Points}: Each dataset execution on a specific transformation configuration (e.g., model version, backend version) produces an evaluation set, comprising evaluation points—each linked to a corresponding datapoint.
    \item[Evaluation Results:] Contains metrics and errors generated by the evaluation tools for each evaluation point.
    \item[Coverage Data:] Defines the coverage model and maps datapoints to the coverage events they address, using a many-to-many relationship.
\end{description}

\begin{figure}[t]
\centering
\includegraphics[width=\columnwidth]{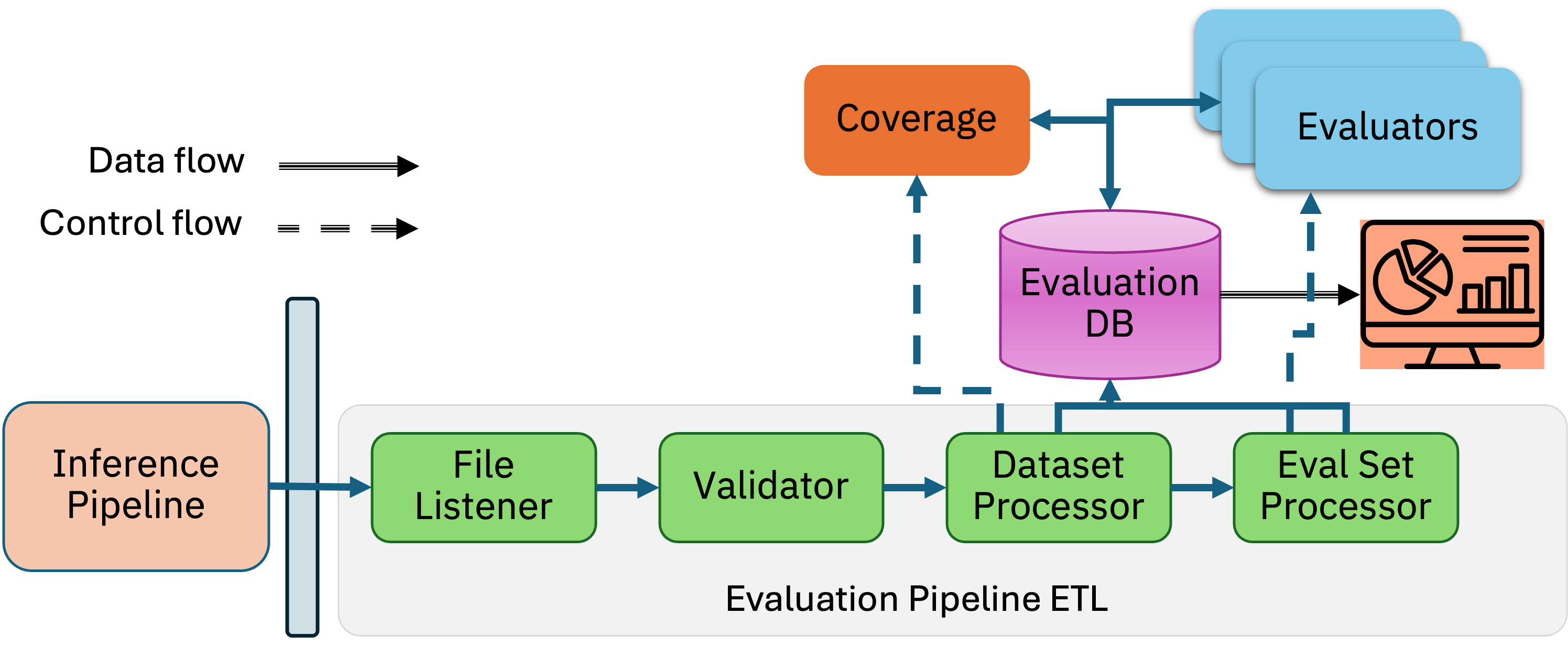}
\caption{Architecture of the evaluation pipeline}
\label{fig:eval-arch}
\end{figure}

Figure~\ref{fig:eval-arch} illustrates the architecture and flow of the evaluation pipeline. The pipeline implements an ETL (Extract Transform Load)~cite{etl}. The processing begins with a file listener that monitors the shared Git repository. When a new \texttt{.jsonl} file (representing an evaluation set) is detected, the listener retrieves and validates it. If valid, the evaluation points are extracted. If the file introduces a new dataset, the dataset and its datapoints are uploaded to the database, and coverage is computed and stored.

Subsequently, the evaluation set and points are stored in the database, and the pipeline triggers the relevant evaluation tools. Each tool retrieves its inputs from the database, performs its analysis, and writes back the resulting metrics and errors. Once all tools complete their tasks, the data becomes available for analysis and reporting.

\section{Evaluation of a Single Translation}
\label{single-trans-sec}

At the core of the evaluation system lies the assessment of a single translation performed by the translating LLM subsystem. This subsystem consists of the LLM itself and a postprocessor that extracts the code from the LLM's output and applies simple transformations to it. The input to the LLM (the prompt) includes not only the source COBOL code but also additional information required for the translation, as shown in Figure~\ref{fig:ex-var-map}. The expected output is a single Java method with the specified signature that replicates the functionality of the source COBOL code.

The goal of the translation evaluator is not merely to determine whether the translation is correct. Rather, it should identify issues in the translated code and estimate the human effort required to correct them.

It is often said that there are no silver bullets in quality evaluation in general, and in software quality evaluation in particular~\cite{brooks1987nosilverbullet}. Before presenting our approach to translation evaluation, we examine several potential ``silver bullets'' and explain why none of them fully address the problem.

\begin{description}
\item[Formal equivalence proof.] The ideal solution would be to mathematically prove that the source COBOL code and the resulting Java code are equivalent. Unfortunately, the code equivalence problem is undecidable~\cite{sipser2012introduction}. While tools such as KLEE~\cite{klee} and CBMC~\cite{cbmc} can perform equivalence checking in specific scenarios, they are limited in scope and cannot bridge the significant semantic gap between COBOL and Java. Thus, formal equivalence checking is impractical in our context.

\item[Compile and execute.] At the other end of the spectrum lies dynamic testing, which in our case involves compiling and executing the translated code and comparing its output to that of the original. Although dynamic testing cannot prove correctness, it remains the leading technique for evaluating software quality~\cite{forgacs2024modern}. However, it is not a silver bullet for several reasons. Chief among them is the effort required to perform thorough dynamic testing. Since translation occurs at the paragraph/method level, we must either construct a test environment for each translation or integrate all translations into a single environment. The former is prohibitively labor-intensive, while the latter has a low probability of success (e.g., if a program has 10 paragraphs and each is translated with 90\% accuracy, the probability of a fully correct program is only about 34\%). Moreover, testing full programs makes it harder to pinpoint specific issues.

\item[Ground truth.] Comparing the translated Java to a given ground truth is another option for quality evaluation. This option is not practical for two main reasons. First, it requires significant SME effort to  create and maintain (e.g., when the Class Designer changes.) Moreover, there can be more than a single correct translation. This means that comparison to a ground truth may require proving equivalence between two Java methods, which is hard.

\item[Human evaluation.] Subject matter experts (SMEs) can review the source COBOL and the translated Java and produce detailed reports that address all evaluation requirements. However, it is not scalable for large, industry-grade evaluations due to the time and effort involved, and it is more severe for old languages, such as COBOL. This does not mean human evaluation is unimportant. It plays a critical role, particularly in evaluating the evaluators themselves, but it cannot serve as the primary evaluation method.

\item[LLM as a Judge (LaaJ).] LLMs can serve as automated judges of translation quality, potentially replacing SMEs. While LaaJs avoid the scalability issues of human evaluation, they have their own limitations. First, they are generally more error-prone than SMEs. Second, in the specific context of COBOL-to-Java translation, LLMs often lack domain-specific knowledge regarding IBM middleware COBOL, such as CICS, which reduces the reliability and quality of their judgments. As such, while LaaJs are a valuable component of the evaluation process, they are not a silver bullet.
\end{description}

Given the absence of a silver bullet, our approach to evaluation of the translation quality relies on several checkers and a diverse set of metrics. These checkers fall into three categories. First, there are static analytic checkers. These checkers utilize static analysis of the translation. The checkers are accurate (or almost fully accurate) in the sense that if a property they are checking fails, they will detect it and if they detect a failure, it is real.  The second category is dynamic testing. Namely, compile and execute\footnote{Technically speaking, compilation is a static checking. It is bundled with execution because of the way it is implemented in our system.}. The third category are LLMs as a Judge (LaaJs) that can provide a holistic view of the translation quality but are less accurate.

A key principle guiding the development of these checkers is a focus on known problematic behaviors of LLMs. For example, their occasional production of gibberish or hallucinated content. The following subsections describe our checkers and metrics in more detail.

\subsection{Syntactic Checking}
The first level of evaluation involves a set of syntactic checks on the translated Java code. These checks do not require access to the COBOL input or an understanding of the code’s intended functionality, making them simple and highly reliable. The syntactic checks include checks for non-empty LLM output on the one hand, and output that does not contain endlessly repeated text on the other hand. Both check for known failure modes of LLMs.

A second set of syntactic checks is based on parsing of the Java code. It includes, for example, checks that the code is parsable and a check that the code contains at least one executable statement. We use the Tree-sitter~\cite{tree_sitter_repo} Java parser due to its ease of use and its ability to produce meaningful parse trees even when the code contains syntax errors.

In the early stages of the project, these syntactic checkers frequently failed and provided valuable insights into issues with the model and its runtime environment. As the project has matured, such failures have become less common, but the checkers continue to offer useful diagnostics regarding the model’s current state.

\subsection{Semantic Checking}
While verifying that the full semantics of the source COBOL code are preserved in the Java translation is infeasible, it is possible to check whether specific semantic elements are correctly translated. The semantic elements we focus on include variable usage, procedure calls (i.e., COBOL \texttt{PERFORM} statements), and certain middleware calls embedded in IBM COBOL for mainframes, such as CICS, IMS, and SQL.

In essence, the checking process involves traversing the control flow graph (CFG) of the COBOL code, obtained using \textit{IBM Application Discovery and Delivery Intelligence} (ADDI) tool~\cite{addi}, and identifying corresponding elements in the Java parse tree. Note that this type of checking may raise false positives in rare but valid cases where a semantic element does not require translation.

\subsubsection{Variable Access Matching}
For variable access matching, we verify that every variable written to (i.e., defined) in the COBOL code is also written to in the Java code, and that every variable read from (i.e., used) in the COBOL code is similarly read in the Java code. The mapping between COBOL and Java variables is defined by the class designer component of the WCA4Z translation system and is provided to both the LLM and the checker as part of the prompt.

Our matching approach is intentionally loose: we check that each access in COBOL has a corresponding access of the same type (read or write) in Java, without enforcing a strict count match. A tighter check of comparing the exact number of accesses was found to produce too many false positives due to structural differences between COBOL and Java control flows.

\subsubsection{Procedure Invocation Matching}
Certain forms of the COBOL \texttt{PERFORM} statement transfer control to one or more paragraphs, effectively functioning as procedure calls. These should be translated into method calls in Java. As with variable matching, the mapping between COBOL procedure calls and Java methods (including method signatures) is defined by the class designer and included in the prompt.

The checker verifies that the LLM-generated Java code includes the correct method calls with the appropriate parameters. As in the variable case, we apply a loose matching strategy, ensuring that each COBOL call has a corresponding Java call without enforcing strict count equivalence.

\subsubsection{Middleware Call Matching}
IBM COBOL includes special statements for interacting with middleware systems, such as \texttt{EXEC SQL}, \texttt{EXEC CICS}, and specific \texttt{CALL} statements for IMS. These should be translated into specific Java method calls, making them relatively easy to identify in both source and target code.

The checking process proceeds in two steps: In the first step, the checker scans the COBOL code (via the CFG) and the Java code (via the parse tree) to identify middleware calls. For each call, it extracts the call type (e.g., IMS Get Next Transaction, SQL SELECT statement) and relevant parameters. This results in two ordered sequences of middleware calls, one from the COBOL code and one from the Java translation.

In the second step, we use the Needleman–Wunsch algorithm~\cite{needleman_wunsch}, a dynamic programming technique originally developed for aligning biological sequences, to align the two sequences. The alignment results point to elemetns that are correctly translated, elements in the COBOL code that have not been translated, hallucinations in the Java code of elements that do not have a corresponding COBOL element, and elements that have matching call types but mismatching parameters.

Note that this check focuses solely on the presence and alignment of middleware calls. It does not verify setup code or result handling, which may differ significantly between COBOL and Java (e.g., return codes in COBOL vs. exceptions in Java). Unlike variable and procedure matching, this check enforces a strict one-to-one correspondence.

\subsubsection{Hallucinations}
LLMs are known to hallucinate~\cite{li2024dawn}. The semantic checks described above can help detect hallucinated elements in the Java code. For {\em variable access}, any access to a variable that is neither part of the COBOL-to-Java mapping nor declared locally in the Java code is flagged as a hallucination. For {\em procedure invocations}, hallucination detection is more difficult. Java code may legitimately include calls to standard libraries, even if those libraries are not explicitly imported in the prompt. Without deeper semantic analysis, we cannot reliably distinguish between legitimate and hallucinated calls, so we do not report hallucinations in this category. For {\em middleware calls}, the situation is clearer. Any Java middleware call that does not align with a COBOL call is considered a hallucination and is reported as such.

\subsection{Compilation and Execution}
Conceptually, dynamic testing of translated code is straightforward: compile and execute the Java code. In practice, however, this process is complex and required significant R\&D effort. The challenges stem from three main sources: First, compiling and executing the translated code requires access to Z platforms that is configured correctly to the source application (e.g., has all the required middleware installed and their Java libraries available.) Second, compilation and execution require full programs, while the translation is done one method at a time. Finally, high-quality dynamic testing requires high-quality stimuli.
 
We built a framework that can compile and run both COBOL and Java program on Z platforms. The framework creates compilation and execution jobs and schedule them on a target Z machine using the galasa testing application for z/OS~\cite{galasa}. The complication and execution results are then collected from the Z host and stored in the evaluation database for farther analysis.

We currently support three modes of compilation: The first mode is compilation of the classes’ skeletons created by the Class Designer. This mode cannot help in the evaluation of the translated code, but it a prerequisite to the other two modes, because if the skeleton does not compile, the classes with generated code are not going to compile as well. In addition, this mode was able to detect several important problems in the Class Design itself. This mode cannot be executed.

The second mode is compilation with a single generated method injected into the classes’ skeleton instead of its stub. This mode can detect issues in the translation that cannot be detected by the parsing static check, such as improper use of class attributes and methods, but many of these issues are detected by other analytic checkers, such as the variable use. Execution of this mode requires construction of a testing environment for each translation. For the reasons discussed earlier in the section, namely the complexity of constructing many such environments, we are not executing this mode.

The third mode is compilation of a fully injected programs. That is, replacing all the methods’ stubs with generated Java methods. With this mode, the Java programs can be tested in the same way their source COBOL program are tested. So far, we have not reached the state in the project where this mode can efficiently be used. 

So far, dynamic testing has yielded the lowest return on investment (ROI) among our evaluation techniques. This is largely due to the current state of the project, which does not yet support error-free translation of full programs. Nevertheless, compiling empty classes or classes containing a single translated method has proven useful for uncovering issues—primarily in the class designer component rather than the LLM itself.

\subsection{LLMs as a Judge (LaaJs)}

Using Large Language Models (LLMs) as evaluators for COBOL-to-Java translation offers a complimatry approach to analytic checking. LLMs are capable of understanding both syntactic structures and semantic nuances across source and target languages. This enables them to assess translation quality more holistically, considering not only syntactic correctness but also semantic fidelity, logic preservation, and adherence to idiomatic programming practices. Analytic systems, in contrast, require exhaustive rule definitions and manual updates. On the other hand, analytic checker are accurate, while LLMs suffer from many inherit issues that affect their reliability.

To construct a Large Language Model as a Judge (LaaJ), the first step involves defining the specific criterion it is intended to evaluate. In this context, the objective is to assess whether the Java translation faithfully preserves the functional semantics of the original COBOL code. Central to this process is the establishment of an evaluation scale that enables consistent and meaningful judgments. We adopted a seven-point scale, developed in collaboration with domain experts who conducted human evaluations of the translation model. This scale, detailed in Table~\ref{tab:scale}, formed the basis of the initial prompt provided to the LaaJ, which combined these descriptive levels with an explicit request to evaluate the correctness and fidelity of the Java translation relative to the COBOL source.

\begin{table*}[ht]
\centering
\caption{Seven-point evaluation scale for assessing COBOL-to-Java translation quality}
\label{tab:scale}
\begin{tabular}{@{}cl@{}}
\toprule
\textbf{Score} & \textbf{Description} \\ \midrule
\textbf{1} & No attempt at translation; output lacks any meaningful correspondence to the source. \\
\textbf{2} & Attempted translation with entirely fabricated classes or methods; functionally non-equivalent and not correctable. \\
\textbf{3} & Partial elements of correct translation present; major errors or hallucinations; fixable with major developer effort. \\
\textbf{4} & Mostly correct translation; moderate errors or hallucinations; fixable with moderate developer effort. \\
\textbf{5} & Mostly accurate translation; minor errors or hallucinations; fixable with minimal developer effort. \\
\textbf{6} & Functionally equivalent translation with verbosity, non-idiomatic constructs, or harmless hallucinations; refinement needed. \\
\textbf{7} & Fully accurate, functionally equivalent, concise, and idiomatic translation. \\ \bottomrule
\end{tabular}
\end{table*}

We employ two complementary methods to validate and refine the prompt design for the LLM as a Judge (LaaJ). The first method is based on human evaluation. Specifically, domain experts assessed a subset of the model outputs, providing both numerical scores and detailed reasoning for their judgments. These annotations serve as a reference for aligning the LaaJ’s output with human evaluators’ expectations. Using the reasoning provided by experts, we iteratively adjust the prompt design to improve consistency between the LaaJ-generated scores and reasoning and human evaluations, while limiting the overfitting.

The second method is based on partial order benchmarks. A key role of the LaaJ within the evaluation lifecycle is to support comparative assessments between translation model versions. Specifically, it enables determining whether a new model iteration produces higher-quality translations than an established baseline. To facilitate this, we developed an automated and scalable benchmark framework for the LaaJ, incorporating explicit expectations regarding relative translation quality into the benchmark dataset.

For each benchmark sample, we generate three variants representing decreasing levels of translation quality. These variants define the expected ordering: Sample \(A\) is anticipated to achieve the highest quality score, Sample \(B\) an equal or lower score, and Sample \(C\) the lowest score. 

At the core of this approach lies the assumption that a well-calibrated LaaJ will align strongly with these expected orderings. Conversely, low alignment scores indicate either deficiencies in the LaaJ’s evaluation consistency or flaws in the benchmark expectations themselves---necessitating refinement of the benchmark artifacts.

Figure~\ref{fig:laaj-prompt} shows the prompt template of the LaaJ we use in the evaluation.
Note that the score explanation is omitted from the prompt because they are identical to the scale presented in Table~\ref{tab:scale}. The prompt contains three main parts: general instructions on the task at hand, details on what and how to score, and specific instructions on pitfalls to avoid. The latter is the result of human analysis of LaaJ scores and reasoning that identified false error detection.

\begin{figure*}[t]
\centering
\begin{tcolorbox}[mypromptbox]
Here is a COBOL program and its Java code translation.
The COBOL program includes the source code, along with variable and class mappings used in the 
translation process.

Please assign a score to the following: correctness of the control flow in the Java code translation.
Use a scale from 1 to 7, where: \\ \noindent

\textit{(Scores omitted. See Table~\ref{tab:scale})} \\ \noindent

For each score, write the reasoning behind the score.
When you find something wrong, explain the problem in detail.

Give the score for the overall Java translation.

If the COBOL program contains an EXIT statement, please ignore it. It should not be translated to Java. 
Do not penalize the score if the EXIT statement was omitted in the Java code.
If the Java code contains any TODO comments, ignore them.
When writing the reasoning, start with \#\#\#Reasoning and end with \#\#\#End\_Reasoning.

If you find any Java translation that has no source in the COBOL code, count this as a hallucination.
Please report the total number of hallucinations in the Java code.

\textbf{COBOL:}
\{COBOL\_code\}

\textbf{Java:}
\{Java\_code\}

\end{tcolorbox}
\caption{Prompt template for COBOL-to-Java translation LaaJ}
\label{fig:laaj-prompt}
\end{figure*} 

\subsection{Checking the running example}

To demonstrate the operation of the checkers, the Java code in Figure~\ref{fig:ex-java-code} contains three injected faults. First, the access to \texttt{CA-CUSTOMER-NUM} COBOL variable in Line~11 of the Java code is not done according to the variable mapping in Figure~\ref{fig:ex-var-map}. Second, the exception in Line~12 is of type \texttt{CicsException} instead of \texttt{CicsConditionalException}. Finally, the call to the \texttt{abend} method in Line~17 uses the wrong value to the dump parameter. 

Table~\ref{tab:ex-errors} shows the errors reported by the analytic checkers and the score and reasoning of the LaaJ. The syntactic checkers all passed and did not report any error. The first two errors are reported by the variable matching checker. They correspond to the first injected fault. The first error indicate that \texttt{CA-CUSTOMER-NUM} is not used in the Java code because of the wrong translation. \texttt{caCustomerNum} that is used instead of the correct access is not declared in the Java code, as the second error indicates. Note that the second error is also detected by the Java compiler because \texttt{caCustomerNum} is not declared in the method or the class. The compiler log is not shown in the figure because of its length. The CICS mismatch in the third error reported by the middleware call matching checker corresponds to the wrong parameter value in the \texttt{abend} method call. Note that the second fault, the wrong exception type, is not detected by the analytic checkers.

\begin{table*}[t]
    \centering
    \caption{Running example evaluation errors}
    \begin{tabular}{|l|c|p{12cm}|}
        \hline
        Type & Score & Message \\
        \hline
        A-VAR & & Variable CA-CUSTOMER-NUM (dfhcommarea.caCustomerNum) is used in the COBOL code but not in the Java code \\
        \hline
        A-VAR & & variable caCustomerNum is not defined in the method or variable mapping \\
        \hline
        A-CICS & & Found mismatch between Statement CICS ABEND and Java location 16 of type ABEND \\
        \hline
        LaaJ & 5 & 
        The score of 5 is given because the Java code translation is mostly correct, but there are some minor errors and differences in the control flow. The issues are: 
        \begin{enumerate}
            \item The `wsResp` and `wsResp2` variables are not initialized before the try-catch block, but they are used in the catch block. In the COBOL code, these variables are not initialized either, but they are assigned values only when an error occurs. 
            \item The Java code catches a `CicsException`, but the COBOL code checks the `WS-RESP` variable to determine if an error occurred. The Java code should catch a `CicsConditionException` instead. 
            \item The `Task.getTask().abend("LGV0", true)` statement is equivalent to the `EXEC CICS ABEND` statement in the COBOL code. 
            \item The `return` statement at the end of the catch block is equivalent to the `EXEC CICS RETURN` statement in the COBOL code. 
        \end{enumerate}
        \\
        \hline
    \end{tabular}
    \label{tab:ex-errors}
\end{table*}

The LaaJ gave the translation a score of 5, indicating that the translation is mostly accurate. In the four items it raised in its reasoning, the first points to a non-existing problem of using variables before they are assigned. The second point is a correct identification of the second injected fault. The third point wrongly claims that the CICS ABEND statement is translated correctly. The last point correctly state that the translation of CICS RETURN is correct. Note that the LaaJ did not detect the first and third injected faults.

Overall, the example shows that none of the checking techniques we use is perfect, a combination of the analytic checkers and LaaJ provides a good assessment about the quality of the translation.

\section{Analysis and Reporting} 
\label{reporting-sec}

The ultimate goal of the evaluation process is to provide a diverse set of users with clear and concise information tailored to their roles. To support this objective, we developed a comprehensive, multi-level analysis and reporting system based on the Grafana platform~\cite{grafana}. Different users and stakeholders require different types of information at varying levels of abstraction, and our system is designed to meet virtually all of these needs.

\begin{figure*}[t] 
\centering 
\includegraphics[width=\textwidth]{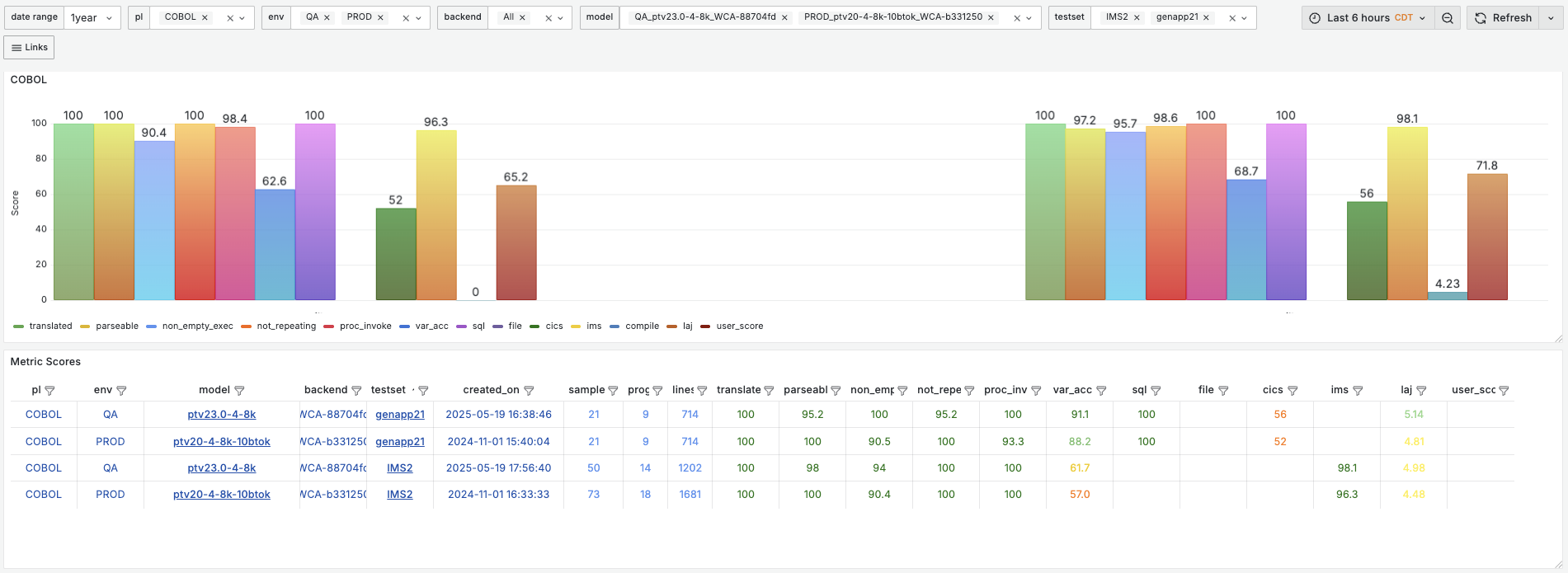} 
\caption{High-level comparison between \texttt{ptv23} and \texttt{wca4z23}} 
\label{fig:high-level} 
\end{figure*}

Project managers, for example, need a high-level overview of project quality and the ability to compare different versions, for example, with different LLMs. Figure~\ref{fig:high-level} presents a high-level comparison between two LLMs, \texttt{wca4z23} and \texttt{ptv23}, evaluated on the genapp21 and IMS2 benchmarks. The bar chart at the top shows the overall scores for each metric, with \texttt{ptv23} on the left and \texttt{wca4z23} on the right. The table below provides the same information broken down by benchmark. The figure illustrates that both models perform similarly in syntactic checks, but \texttt{ptv23} outperforms \texttt{wca4z23} in middleware semantic checks. For LaaJ scores, \texttt{ptv23} performs slightly better on IMS2 but worse on genapp21.

Based on the report in Figure~\ref{fig:high-level}, a project manager might decide to investigate the inconsistency between semantic checks and LaaJ scores in genapp21. The manager assigns this task to a technical lead, who then examines the scores for all evaluation points in the benchmark using the {\em all samples view}. The lead selects a few noteworthy evaluation points, such as those with high LaaJ scores but few semantic errors (as in the running example), and assigns them to team members for deeper analysis.

The assigned team member uses the {\em single sample debug view}, which provides all necessary information for debugging. This includes the original COBOL source code (Figure~\ref{fig:ex-cobol-src}), the translated Java code (Figure~\ref{fig:ex-java-code}), the variable mapping (Figure~\ref{fig:ex-var-map}), a table of checker scores, and the reported errors (Table~\ref{tab:ex-errors}). The team member can also utilize views that compare two translations and show the differences between them.

\begin{figure}[t] 
\centering 
\includegraphics[width=\columnwidth]{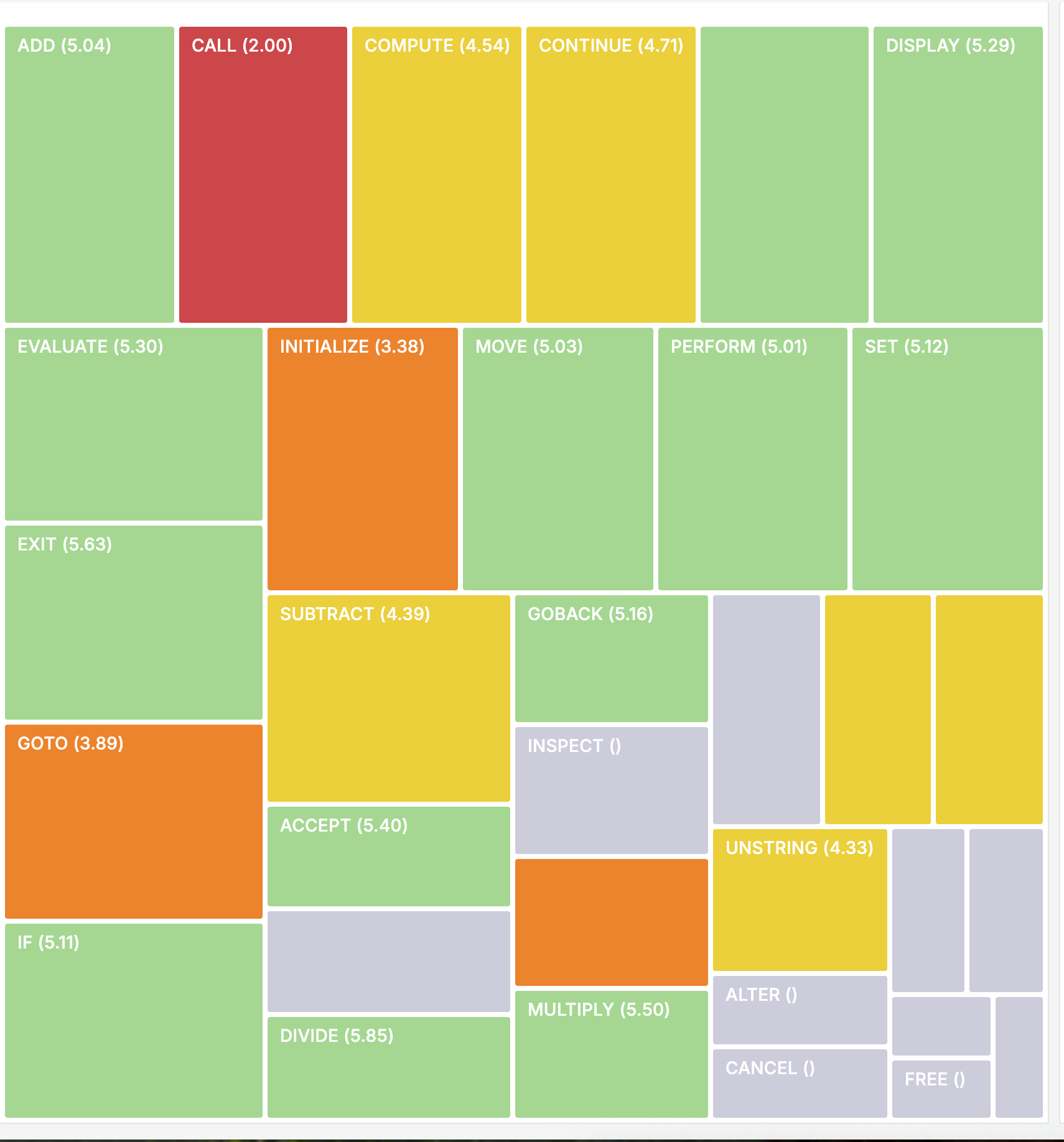} 
\caption{Heatmap for COBOL statements} 
\label{fig:heatmap} 
\end{figure}

The reports and views described so far are based on straightforward summarization of checker results stored in the database. However, our analysis and reporting system also supports deeper insights. For instance, when a team lead wants to identify problematic areas in the subsystem performance, they can use the collected coverage data to explore correlations between COBOL features (e.g., specific statements or middleware transactions) and translation quality (e.g., reflected in LaaJ scores). Figure~\ref{fig:heatmap} shows a heatmap of LaaJ scores by COBOL statement. Each cell represents a COBOL statement and displays the weighted average LaaJ score across all evaluation points in the selected benchmarks containing that statement. Gray cells indicate statements not present in any evaluation point in the benchmarks. For the rest, the color and score reflect the average translation quality. For example, the ADD statement shows a relatively high score of 5.04, while CALL has a lower score of 2, indicating a need for improvement.

A simpler coverage-based report presents a hierarchical view of our coverage model. At each level of the hierarchy, the report shows how frequently each element is covered. For example, at the subcategory level, the report may show that the INSPECT statement is not covered at all, ADD is covered nearly 100 times, and CALL is covered only once. This report helps the benchmark authoring team identify areas that require additional coverage or enhancement.

\balance
\section{Summary and Future Work} 
\label{summary-sec}

In this paper, we presented the quality evaluation framework for the COBOL-to-Java code translation subsystem of IBM watsonx Code Assistant for Z (WCA4Z). The evaluation platform is a data-centric system that receives translated code from the product, processes it, and applies a variety of checkers and evaluators. The results are stored in a database and used for analysis and reporting, ranging from detailed inspection of individual translations for debugging purposes to high-level quality assessments and comparisons across different versions of the subsystem.

The platform has been in use since the early stages of the WCA4Z project and has matured alongside it. It has played a critical role in improving the quality of the translation component by offering a clear and consistent view of overall performance. This visibility has helped build management’s confidence in the system. Its analytical capabilities and support for deep dives have enabled the identification of significant issues and the resolution of performance gaps. Furthermore, the platform’s foundational components, including the object model, processing pipeline, database, and several checkers, have been adapted for evaluating other WCA4Z components, such as code explanation and code generation.

Despite its effectiveness, the evaluation platform and the assessment of individual translations are still evolving. We continue to enhance them in several directions. In the area of semantic analysis, we are exploring methods to bridge the semantic gaps between COBOL and Java, such as differences in CICS transaction error handling, as illustrated in Figure~\ref{fig:running-ex}.

To further improve LaaJ’s evaluative capabilities, we are investigating techniques for incorporating domain-specific knowledge that is not inherently available in out-of-the-box language models. This includes addressing limitations in understanding specialized COBOL and Java constructs, legacy system behaviors, and industry-specific conventions. By integrating such knowledge, through prompt engineering, fine-tuning with domain-relevant datasets, or hybrid approaches that combine symbolic rules with LLM reasoning, we aim to achieve deeper semantic alignment with expert evaluations and more contextually accurate assessments.

Finally, we are also working on better identifying problematic areas in translation quality through in-depth analysis of evaluation results and field feedback. Our goal is to use these insights to automatically enhance the evaluation process by refining our coverage models and generating additional benchmarks that target these challenging areas.

\clearpage
\bibliographystyle{IEEEtran}
\bibliography{reference}

\end{document}